\documentclass[12pts,twocolumn,epsf,aps,prb]{revtex4}
\usepackage{color,graphicx}
\usepackage{hyperref}
\begin{document}

\title{Critical exponents and irreversibility lines of La$_{0.9}$Sr$_{0.1}$CoO$_3$ single crystal}

\author{N. Khan$^1$, A. Midya$^1$, P. Mandal$^1$, and D. Prabhakaran$^3$}

\affiliation{$^1$Saha Institute of Nuclear Physics, 1/AF Bidhannagar, Calcutta 700 064, India}
\affiliation{$^3$Clarendon Laboratory, Department of Physics, University of Oxford, Oxford OX1 3PU, United Kingdom}

\date{\today}
\begin{abstract}

We have studied the dynamic and static critical behavior of spin glass transition in insulating La$_{0.9}$Sr$_{0.1}$CoO$_3$ single crystal by ac susceptibility and dc magnetization measurements in the vicinity of its freezing temperature ($T_f$). The dynamic scaling analysis of the frequency dependence of ac
susceptibility data yields the characteristic time constant $\tau_{0}$=1.6(9)$\times$10$^{-12}$ s, the dynamic critical exponent $z\nu$=9.5(2), and a frequency
dependence factor $K$=$\Delta$$T_{f}$/$T_{f}$($\Delta$$log$$f$)=0.017, indicating that the sample enters into a canonical spin-glass phase below $T_{f}$=34.8(2)
K. The scaling analysis of non-linear magnetization in the vicinity of $T_{f}$ through the static scaling hypothesis yields critical exponents $\beta$=0.89(1)
and $\gamma$=2.9(1), which match well with that observed for well known three-dimensional (3D) Heisenberg spin glasses. From the longitudinal component of
zero-field-cooled and field-cooled magnetization measurement we have constructed the $H-T$ phase diagram which represents the field evolution of two
characteristic temperatures: the upper one, $T_{w}(H)$, indicates the onset of spin freezing in a uniform external field $H$, while the lower one, $T_{s}(H)$,
marks the onset of strong irreversibility of the frozen state. The low field $T_{s}(H)$ follows the critical line suggested by  d'Almeida-Thouless model for
canonical spin glass, whereas the $T_{w}(H)$ exhibits a re-entrant behavior with a maximum in the $T_{w}(H)$ at a nonzero field above which it follows the
Gabay-Toulouse (GT) critical line which is a characteristic of Heisenberg spin glass. The reentrant behavior of the GT line resembles that predicted
theoretically for $n$-component vector spin glasses in the presence of a uniaxial anisotropy field.    \\

{PACS number(s):75.40.Cx,75.47.Gk,71.30.+h}
\end{abstract}
\maketitle
\newpage

\section{INTRODUCTION}
Though the  perovskite cobaltites don't exhibit colossal magnetoresistance and have low ferromagnetic Curie temperatures relative to their manganite counterparts, these materials received much attention due to a couple of unique properties; namely, the presence of spin-state transition of the Co ion and unusual magnetic ground state.\cite{Rodriguez,Okajima,itoh,nam,hnam,phuc,Yamaguchi,zobel,korotin,mandal} Among the doped perovskite cobaltites, the prototype large bandwidth La$_{1-x}$Sr$_{x}$CoO$_3$ (LSCO) system has been studied extensively and is also a model system for studying the magnetoelectronic phase separation (MEPS) phenomenon refers to the spatial coexistence of multiple electronic and magnetic phases on nanoscopic length scale.
\cite{raccah,senaris,rcaciuffo,caciuffo,kuhns,rhoch,hoch,jwu,dphelan,he,che,locua,phelan} The MEPS is believed to be responsible for the intriguing physical
properties of complex transition metal oxides like cuprates and manganites.\cite{dagotto,edagotto} The undoped LaCoO$_{3}$ is a non-ferromagnetic insulator.\cite{mitoh} With the substitution of Sr$^{2+}$ at La$^{3+}$ site in LaCoO$_3$, the system phase separates into nanoscopic hole-rich ferromagnetic (FM) metallic clusters  in a hole-poor insulating matrix. The ferromagnetic double exchange interactions between Co$^{4+}$ and Co$^{3+}$ in hole-rich clusters coexist with antiferromagnetic (AF) superexchange interactions between Co ions with same valance states (Co$^{3+}$$-$Co$^{3+}$ and Co$^{4+}$$-$Co$^{4+}$).\cite{senaris,bhide,ganguly} The competition between the coexisting FM and AF interactions together with randomness which results from the distribution of different valance Co ions in the crystal gives rise frustration that leads to a spin glass (SG) state for 0.05$\leq$$x$$\leq$0.18.\cite{itoh,mitoh,wu} As the Sr concentration increases, the FM metallic clusters grow in size and numbers and percolate through the non-FM matrix, leading to a transition from insulating spin glass phase to FM metallic state at a percolation threshold $x_p$$\approx$0.18.\cite{jwu,dphelan,Aarbogh} Recent studies on the size of magnetic cluster and the FM phase fraction in LSCO single crystals reveal that the MEPS is confined to a well-defined doping range, 0.04$<$$x$$<$0.22.\cite{he,che} But in polycrystalline samples this MEPS may exist even upto $x$=0.5 due to several extrinsic effects.\cite{kuhns,rhoch,hoch} In order to know the exact nature of the magnetic ground state, extensive investigations were carried out using both local and bulk experimental techniques in this system over a wide range of doping, particularly in the FM side of its phase diagram using single crystalline materials. To be mentioned, for single crystals with 0.18$<$$x$$<$0.22, the system exhibits a true long-range ferromagnetic ordering but the FM phase fraction is less than 100\%.\cite{he,che} Samples with $x$$>$0.22 exhibit magnetic features similar to that of  conventional homogeneous FMs.\cite{he,che} Recently, we have also shown this phase separation scenario around $x$=0.22 by studying the critical behavior of FM to PM phase transition for the $x$=0.21, 0.25, and 0.33 single crystals where it was observed that the $x$=0.25 and 0.33 compounds behave as 3D Heisenberg FMs with near neighbor interactions, but for the $x$=0.21 compound the behavior deviates from 3D Heisenberg towards mean-field-like due to the presence of magnetoelectronic phase inhomogeneity.\cite{khan,nkhan} Unlike single crystals, the ploycrystalline samples exhibit glassy ferromagnetic behavior over the entire doping range 0.18$<$$x$$<$0.50.\cite{itoh,nam,wu,mukherjee}\\

Relatively less attention was paid in the low doping regime (0$<$$x$$<$0.18) as far as magnetic ground state is concerned and particularly in single crystals.
The existence of spin glass phase in the LSCO system for 0.05$\leq$$x$$\leq$0.18 was shown by some previous reports based on the observed characteristics of spin glass behavior like an extremely slow spin dynamics, ageing, memory effects, and bifurcation in the ``zero-field-cooled" and ``field-cooled"  low field
thermomagnetic curves.\cite{itoh,mitoh,hnam,wu,ktang} The dynamic scaling analysis in the polycrystalline $x$=0.15 sample revealed that there is no true spin
glass phase at finite temperatures\cite{mira} whereas neutron scattering study on the La$_{0.92}$Sr$_{0.08}$CoO$_{3}$ single crystal suggested that the sample
enters into a spin glass phase at low temperatures despite the presence of short-range FM correlation at all temperatures below the onset of
freezing.\cite{kasai} Recent studies in single-crystalline samples based on ac- susceptibility or ageing effects have also shown that the low temperature spin
glass phase exits for the $x$=0.15 and 0.18 compounds.\cite{tang,kmanna,manna} It was also observed that in some polycrystalline samples the ZFC and FC magnetization curves bifurcate well above the freezing temperature whereas in the case of single crystals the bifurcation occurs very close to the
transition.\cite{wu,kmanna,mandal} Several magnetic properties in low doped LSCO system are not determined unambiguously because it is extremely difficult to
achieve a full homogeneous distribution of Sr$^{2+}$ ions in polycrystalline samples prepared using standard solid state reaction techniques. The ``ferromagnetic cluster" effects arising from the inhomogeneous distribution of Sr$^{2+}$ ions in polycrystalline samples are responsible for the ambiguous magnetic properties.
The above mentioned characteristics of the frozen low temperature phase are not exclusive to the ``atomic" or canonical spin glasses. Low doping samples exhibit
frequency dependent ac susceptibility peaks and time dependent phenomena characteristics of a spin glass but also show large FC magnetization and irreversibility temperature indicating the presence of strong FM correlations.\cite{wu} Ample evidences exist in the literature where such glassy behaviors have also been assigned in cluster glasses\cite{nam,dsamal,rivadulla} and superspin glasses.\cite{sahoo} Though these observed behaviors are consistent with a spin glass transition, but could not be taken as a definite proof of it as the freezing is not well understood in terms of thermodynamic phase transition. The standard theoretical model of spin glasses considers Heisenberg interactions with random exchange coupling between the spins in the presence or absence of weak Ising anisotropy and thereby leads to 3D Ising or Heisenberg universality class.\cite{viet,kawamura,hukushima,hasenbusch,katzgraber} So, in order to establish a true spin glass phase and to assign a universality class or for a meaningful comparison between theory and experiments a thermodynamic characterization of the SG transition in this LSCO system should be performed in a high quality single-crystalline sample.\\

The modification of the spin glass transition and the behavior of the frozen state in a uniform external field have also received much attention both
theoretically and experimentally.\cite{edwards,sherrington,almeida,gabay,parisi,vieira,chamberlin,maksimov,kenning,banavar,lago} Whether the transition is a true thermodynamic phase transition is still a subject of controversy. Two main rival theories are the replica-symmetry breaking (RSB) theory of Parisi,\cite{parisi}
which is motivated by the exact solution of Sherington-Krikpatrik (SK) mean-field model\cite{sherrington} and the droplet-scaling
theory.\cite{fisher,bray,mcmillan} The striking result of the SK model is the existence of a phase transition in the presence of a uniform magnetic field
signaled by the d'Almeida-Thouless\cite{almeida} and Gabay-Toulouse\cite{gabay} critical lines in the $H-T$ phase diagram whereas according to the droplet model
there is no phase transition as the spin glass phase is destroyed by any finite applied field. Experimentally, it was observed that $H-T$ phase diagrams obtained for numerous real spin glasses\cite{kenning} are qualitatively similar to that predicted by the mean-field SK model even though most are closer to the 3D
Heisenberg Edwards-Anderson (EA) model, for which numerical simulations predict a zero-temperature transition already in zero field.\cite{banavar} So the LSCO
system will also be a good candidate for studying the spin glass behavior in the presence of a uniform external magnetic field.\\

Here we present a detailed study of dynamic and static scaling analysis for a high quality La$_{0.9}$Sr$_{0.1}$CoO$_3$ single crystal. To avoid the
possibility of dynamic blocking of the interacting spin clusters, we have chosen $x$=0.1 which is well below the percolation threshold ($x_{p}$) where the ``FM cluster'' effects dominates i.e., $x$ lies deep inside the so-called spin glass region of its phase diagram.\cite{itoh,wu} The dynamic scaling analysis shows a canonical spin glass phase below its freezing temperature and the scaling of static nonlinear magnetization shows 3D Heisenbergs spin glass behavior for the compound. The obtained $H-T$ phase diagram where the irreversibility lines maps the different stability regions of a spin glass system is consistent with that predicted for the 3D Heisenberg spin glasses in the presence of a uniaxial anisotropy field.\cite{vieira}
 \\

\section{EXPERIMENTAL DETAILS}

The polycrystalline homogeneous powder sample La$_{0.9}$Sr$_{0.1}$CoO$_3$ was prepared by the conventional solid state reaction method using high purity and
preheated La$_2$O$_3$, SrCO$_3$, and Co$_3$O$_4$ which were mixed in appropriate ratios and sintered in air at 1000$-$1100 $^{\circ}$C for few days with
intermediate grindings. The powder sample was then pressed into cylindrical rods which were finally sintered at 1200 $^{\circ}$C for 24 h in vertical sintering
furnace. Single crystal was grown from these polycrystalline rods by the traveling solvent float zone method using a four mirror image furnace (Crystal System
Inc.) The phase purity and high quality crystalline nature were found from the characterization using different experimental techniques such as x-ray powder
diffraction, Laue diffraction, scanning electron microscopy (SEM) and electron probe microanalysis. The powder diffraction pattern of the crystal can be indexed
by a rhombohedral unit cell with space group $R\bar{3}c$. The thermogravimetric analysis (TGA) using a Rheometric Scientific STA1500 instrument verified that the oxygen content in the sample is close to 3. The details of synthesis, characterization and structural analysis have been reported elsewhere.\cite{mandal} The ac
susceptibility measurements were performed using superconducting quantum interference device magnetometer (Quantum Design). The dc magnetization measurements
were done with a Quantum Design (QD) Physical Properties Measurement System (PPMS) equiped with a vibrating sample magnetometer (VSM). The sample was first
cooled to the lowest temperature 2 K, which is well below the freezing temperature $T_{f}$, then the measurement field was applied and set in the persistent mode and the  zero-field-cooled magnetization were recorded with increasing temperature at the rate of 8 mK/s to minimize the effect of any thermal lag between the
sensor and the sample.\cite{chamberlin} After completing the ZFC measurement, the temperature was again reduced to low temperature without changing the
measurement field, then increased again at the same rate and field cooled magnetization was recorded. The ZFC and FC magnetization were taken at several
measuring fields in the range 50$\leq$$H$$\leq$7000 Oe using the protocol described above. The dc magnetic isotherms in the applied field up to 3500 Oe were
measured in the vicinity of $T_{f}$. Prior to measurements, the sample was cooled down from 300 K $-$ well above the freezing temperature $T_{f}$, to the
prescribed measuring temperature in the absence of field. After stabilizing each temperature another 600 s wait time was given to allow the sample to reach thermal equilibrium before taking the field-dependent data with each field in the persistent mode and then the sample was demagnetized following standard procedure for taking the next isotherm using the same procedure.        \\

\section{Results and Discussion}

\begin{figure}
  \includegraphics[width=0.5\textwidth]{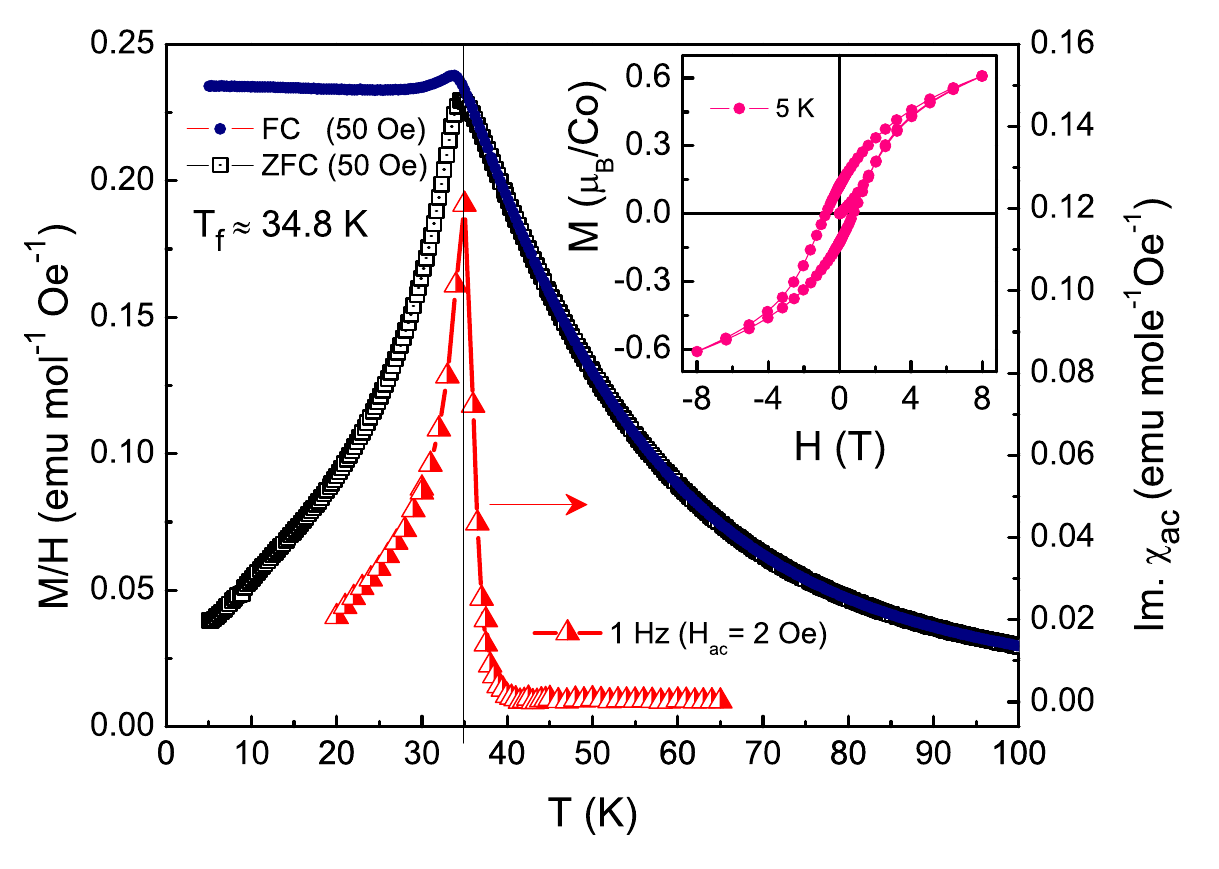}\\
\caption{(Color online) Temperature dependence of the field-cooled and zero-field-cooled dc magnetization at 50 Oe and the imaginary part of ac susceptibility for excitation frequency of 1 Hz and rms amplitude $H_{ac}$ of 2 Oe for the La$_{0.9}$Sr$_{0.1}$CoO$_{3}$ single crystal. The inset shows the field dependence of the dc magnetization at 5 K.}\label{FIG. 1}
\end{figure}

Figure 1 depicts the temperature dependence of dc susceptibility of the La$_{0.9}$Sr$_{0.1}$CoO$_3$ single crystal in an applied field of 50 Oe and the imaginary part of ac susceptibility $\chi^{''}$ at an excitation frequency of 1 Hz of the driving ac field. The ZFC curve shows a cusp at about $T_{f}$$\simeq$35 K and a sharp rise in $\chi^{''}$ at the same temperature. The field cooled curve shows a kink just below the freezing temperature and  the FC and ZFC curves coalesce just above the $T_{f}$ reflecting the high crystalline quality of the sample. The inset shows the field dependence of magnetization at 5 K, which shows S-shaped curve with a small hysteresis but no saturation like behavior is observed at magnetic field up to 8 T. These observed behavior of the temperature dependence of susceptibilities as well as the field dependence of magnetization are the characteristics of a conventional spin glass and predicts a freezing temperature $T_{f}$$\simeq$35 K, which is the onset of freezing of Co moments into a spin-glass state. However, these characteristic features can not be taken as a definitive proof of a true spin glass phase below $T_{f}$ as these behaviors may also result from a dynamic blocking of superparamagnetic like clusters,\cite{nam,dsamal,rivadulla,sahoo} particularly in this material where spontaneous magnetoelectronic phase separation occurs due to formation of hole rich FM clusters in non-FM semiconducting matrix. Therefore its dynamic and static magnetic behavior in the vicinity of $T_{f}$ have been investigated and presented in the following sections. \\

\subsection{Spin dynamics: critical slowing down}

\begin{figure}
  \includegraphics[width=0.5\textwidth]{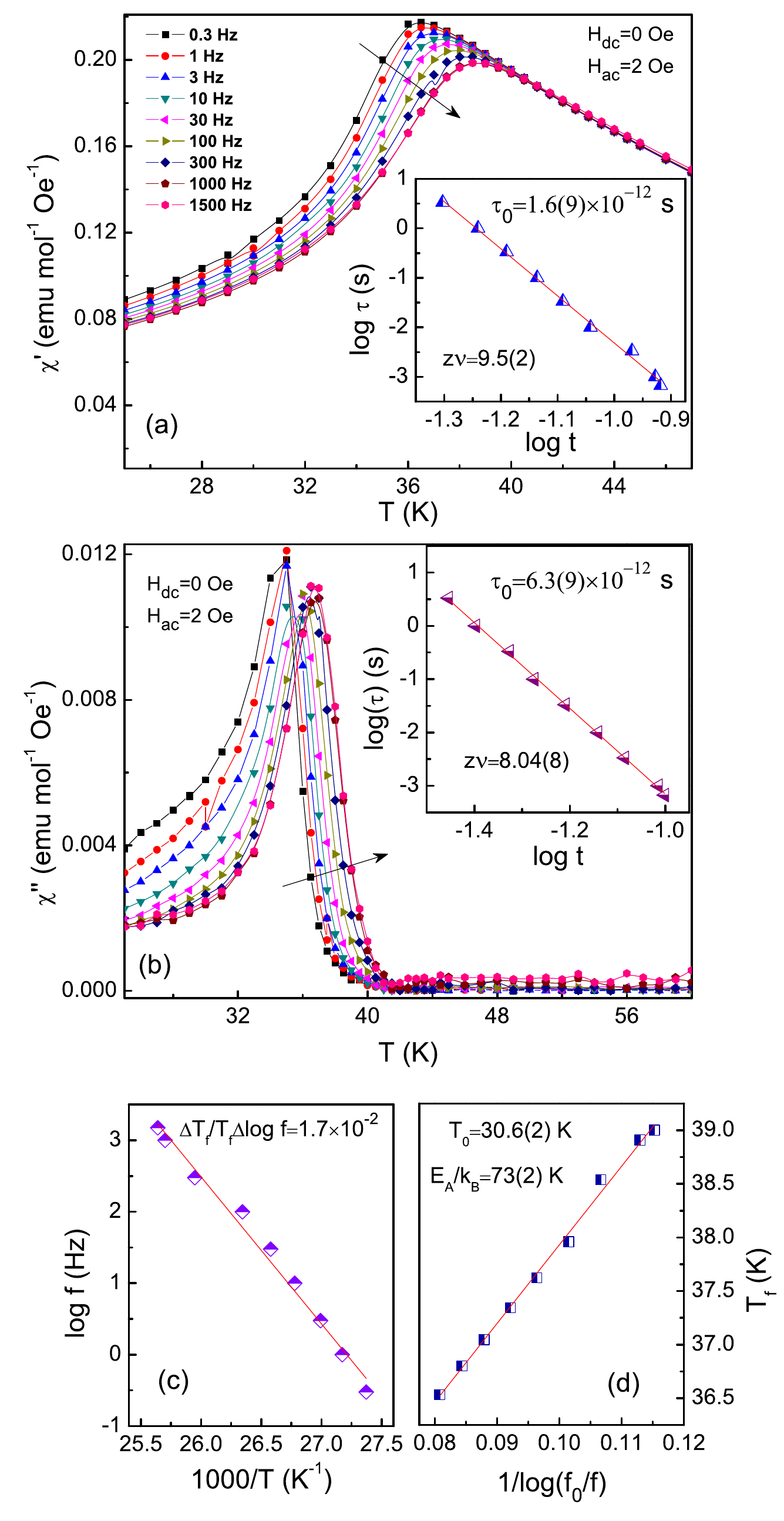}\\
\caption{(Color online) (a) Temperature dependence of the real part of ac susceptibility for different frequencies of the driving ac field where $H_{dc}$=0 Oe and $H_{ac}$=2 Oe. (Inset) Log-log plot of the characteristic time $\tau=1/f$ vs reduced temperature $t$. The solid line is the fit to Eq. (1). (b) Temperature dependence of the out of phase susceptibility at different driving frequencies for the same ac field. (Inset) Log-log plot of $\tau$ vs $t$ and the solid line is fit to Eq. (1). (c) Frequency dependence of the freezing temperature plotted as log $f$ vs 1000/$T$ used to estimate the relative variation by frequency decade ($K$). (d) Vogel-Fulcher representation of frequency dependence of the freezing temperature and the solid line is fit to Eq. (3)}\label{FIG. 2}
\end{figure}

Figure 2(a) shows the temperature dependence of the real part of ac susceptibility ($\chi'$) and Figure 2(b) shows the out of phase susceptibility ($\chi''$) at
different frequencies of the driving ac field. The $\chi'(T)$ displays a cusp-shape, the maximum of which indicates the freezing temperature ($T_{f}$), which
shifts towards higher temperatures with the increasing frequency of the applied ac field.  However, at a fixed frequency, the maximum in $\chi'(T)$ and
$\chi''(T)$ occurs at slightly different temperatures, which was also observed for the $x$=0.15 single crystal.\cite{kmanna} The reason behind this is because of $\chi'(T)$ and $\chi''(T)$ may not be correlated and they might have different origin.\cite{nam} The $\chi''(T)$ represents the magnetic energy dissipation in
the sample and is proportional to the area of the hysteresis loop within one period of the driving ac field at an equilibrium temperature. Therefore, the maximum in $\chi''(T)$ corresponds to the temperature where the hysteresis loop area has an extreme value, which decreases with increasing temperature and becomes zero in the paramagnetic state where there is no hysteresis. So the inflection point in the high temperature side of $\chi''(T)$ reflects the onset of spin freezing and therefore, also gives an estimate of $T_{f}$. For temperature below $T_{f}$, the magnitude of both $\chi'(T)$ and $\chi''(T)$ is frequency dependent and for $T>T_{f}$, the frequency dependence of both $\chi'(T)$ and $\chi''(T)$ almost disappears. All the above features are the well known characteristics of a
conventional spin glass. The relative variation of the freezing temperature $T_{f}$ (determined from maximum of $\chi'(T)$ ) per frequency decade is a
characteristic constant which is defined as $K$=$\Delta$$T_{f}$/$T_{f}$($\Delta$$log$$f$), where $f$ is the frequency of the driving ac field. We have obtained a value of $K$=0.017 using the $\chi'(T)$ data as shown in Figure 2(c). This value of $K$ matches well with that obtained for the $x$=0.15 single crystal\cite{kmanna} and falls within the range reported for canonical spin glass systems and also significantly smaller than the value for superparamagnetic
systems.\cite{mydosh,jltholence} So the above results indicate a slowing down of electron spin dynamics that occurs when the freezing point is approached from
above. Assuming this as a conventional critical slowing down process, the correlation time $\tau$ should therefore, diverge according to $\tau\propto\xi^z$,
where $\xi$ and $z$ are the correlation length and dynamic scaling exponents, respectively.\cite{hohenberg} For a continuous phase transition the correlation
length diverges with temperature as $\xi\propto t^{-\nu}$, where $t$=$(T-T_{f})/T_{f}$ and $\nu$ is the static critical exponent. So, the evolution of the
correlation time with temperature in the vicinity of $T_{f}$ is given by,
\begin{equation}
\tau=\tau_{0}t^{-z\nu},
\end{equation}
where $\tau_{0}$ is the characteristic time of a single spin flip. The inset of Figure 2(a) shows the correlation time $\tau$, which is the reciprocal of frequency of the driving ac field, against the reduced temperature $t$ in a log-log plot where $T_{f}=34.8$ K. A linear fit to the data following Eq. (1) yields $z\nu$=9.5(2) and $\tau_{0}$=1.6(9)$\times$10$^{-12}$ s. The values of the exponent and the characteristic time constant fall well within the range reported for well established canonical spin glass systems.\cite{hohenberg,tien} We have also estimated the value of $z\nu$=8.04(8) and $\tau_{0}$=6.3(9)$\times$10$^{-12}$ s from $\chi^{''}(T)$ data by taking the inflection points as the onset of freezing as shown in the inset of Figure 2(b). It should be noted that though the values of $z\nu$ and $\tau_{0}$ estimated from $\chi^{''}(T)$ are slightly different from that obtained from the $\chi^{'}(T)$ data, their magnitudes are within the realm of conventional spin glass phase.\cite{hohenberg,tien}    \\

Considering the possibility of dynamic blocking of the interacting spin clusters that form in the non-FM matrix, we have also analyzed the spin dynamics by using the empirical Vogel-Fulcher law,\cite{tholence}
\begin{equation}
\tau=1/f=\tau^{'}exp\left(\frac{E_{A}}{k_{B}(T_{f}-T_{0})}\right),
\end{equation}
where $\tau^{'}$ corresponds to $\tau_{0}$ in Eq. (1) and $T_{0}$ is a phenomenological parameter which is often interpreted as the measure of intercluster interaction strength. So, the relative variation of $T_{f}$ with the frequency of driving ac field following Eq. (2) is given by,
\begin{equation}
T_{f}=\frac{E_{A}/k_{B}}{ln(f_{0}/f)}+T_{0},
\end{equation}
The activation energy $E_{A}$ and the parameter $T_{0}$ can be estimated by plotting $T_{f}$ as a function of 1/ln($f_{0}/f$) as shown in the Figure 2(d). A linear fit to the data yields $E_{A}/k_{B}$=73(2) K and $T_{0}$=30.6(2) K. According to the criterion introduced by Tholence,\cite{jltholence} $\alpha$=$(T_{f}-T_{0})/T_{f}$ should be very small for spin glass behavior. We estimate a value of $\alpha$=0.12 which is well within the range of the well-behaved spin glasses.\cite{yeshurun} Also the ratio of the activation energy and $T_{0}$  gives a measure of the strength of interactions between the dynamic entities freezing at $T_{f}$ and hence the level of magnetic clustering where the size of the clusters are assumed to be directly related to the coupling between them.\cite{fiorani} We obtained a value of $E_{A}/k_{B}T_{0}$=2.39(8), which also lies well within the range for canonical spin glass systems
($E_{A}/k_{B}T_{0}$=2$-$3).\cite{fiorani} Therefore, these results are in favor of a true spin glass behavior in single-crystalline La$_{0.9}$Sr$_{0.1}$CoO$_3$.\\

\subsection{Static critical behavior}
\begin{figure}
  \includegraphics[width=0.5\textwidth]{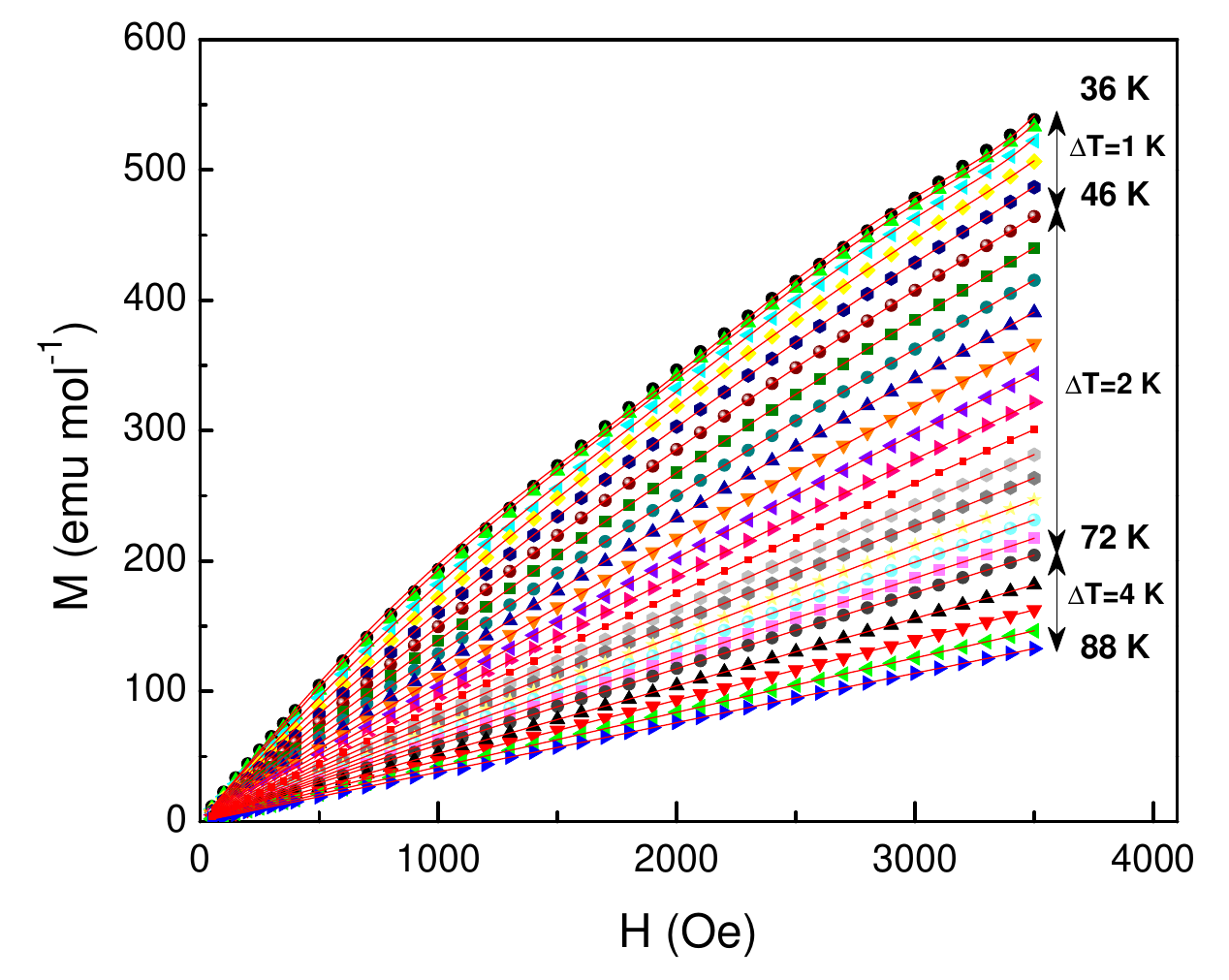}\\
\caption{(Color online) Isothermal magnetization (\emph{M} vs \emph{H}) curves at temperatures in the vicinity of $T_{f}$ in the field range
0$\leq$$H$$\leq$3500 Oe for the La$_{0.9}$Sr$_{0.1}$CoO$_3$ single crystal. The solid lines are due to the fit to Eq. (4).}\label{FIG. 3}
\end{figure}

The above scaling analysis of spin dynamic points towards a true equilibrium spin glass transition around $T_{f}$=34.8 K in the La$_{0.9}$Sr$_{0.1}$CoO$_3$
single crystal. We have therefore studied its static critical behavior in the vicinity of $T_{f}$ following the static critical scaling hypothesis. A
paramagnetic to spin glass phase transition can be characterized by measuring the nonlinear susceptibility $\chi_{nl}$.\cite{kenning,lago} Though the order parameter of a spin glass is still a matter of controversy, it is the squared spatial correlation function $\langle S_{i}S_{j}\rangle^{2}$ that is found to become long ranged below $T_{f}$, resulting in the divergence of the so-called spin glass susceptibility $\chi_{SG}$=$\beta^{2}[(\langle S_{i}S_{j}\rangle-\langle S_{i}\rangle\langle S_{j}\rangle)^{2}]_{av}$.\cite{mydosh,khfisher} This $\chi_{SG}$ is related\cite{suzuki,chalupa} to the $\chi_{nl}$ in such a fashion that, it should also exhibit divergence behavior below $T_{f}$. $\chi_{nl}$  is defined as the higher order contributions in the expansion of the magnetization ($M$) in powers of a uniform external field,
\begin{equation}
M=\chi_{1}H-\chi_{3}H^{3}+\chi_{5}H^{5}-......,
\end{equation}
The expansion does not retain coefficients with even power of $H$ because they are proportional to the spontaneous magnetization $M_{S}$ or its powers, and $M_{S}$=0 in both the spin glass and paramagnetic phases.\cite{suzuki} So the nonlinear susceptibility is given by,
\begin{equation}
\chi_{nl}=\chi_{1}-\frac{M}{H}=\chi_{3}H^2-\chi_{5}H^4+.....,
\end{equation}

\begin{figure}
  \includegraphics[width=0.5\textwidth]{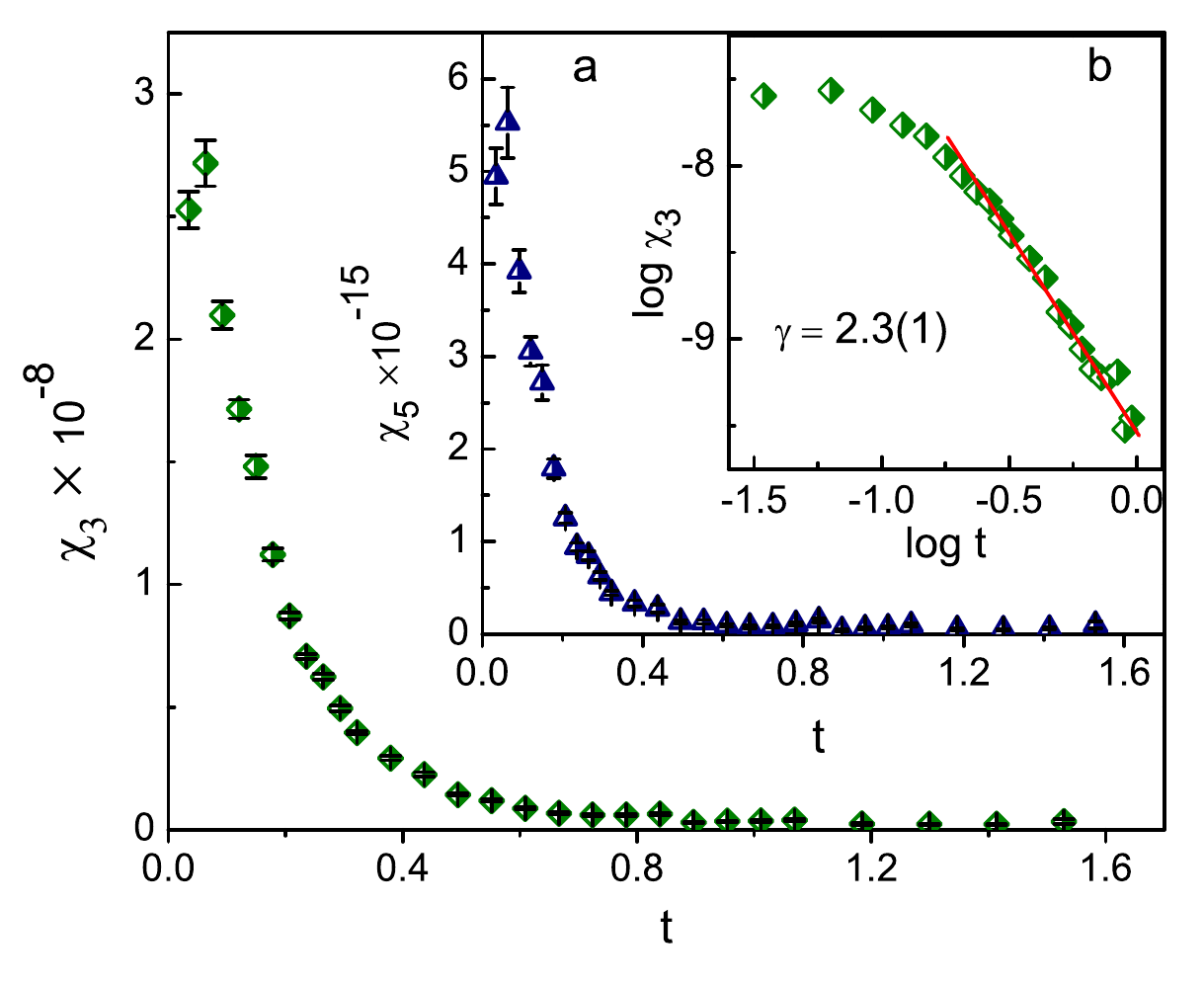}\\
\caption{(Color online) Temperature dependence of the coefficients of the first two nonlinear terms in the expansion of the magnetization i.e., $\chi_{3}$ and $\chi_{5}$ (inset a) plotted against reduced temperature $t$ exhibiting their divergence as $T\rightarrow T_{f}^{+}$. Inset b shows the log-log plot of $\chi_{3}$ versus reduced temperature $t$ and the solid line is due the linear fit for $t\geq0.2$ where $\gamma$ is estimated from the slope of the fit.}\label{FIG. 4}
\end{figure}

Similar to the method performed for the SrFe$_{0.9}$Co$_{0.1}$O$_{3}$ compound by Lago $\emph{et al.}$\cite{lago} to estimate $\chi_{nl}$, unlike ac measurements in the static limit, we have used dc magnetization isotherms in the vicinity of the freezing transition as shown in the Figure 3. The thermal and field evolution of $\chi_{nl}$ in the critical regime can then be analyzed using a universal scaling equation of state\cite{mauger} of the form
\begin{equation}
M_{nl}(t,H)=\chi_{nl}H=t^{(\gamma+3\beta)/2}F\left( H/t^{(\gamma+\beta)/2}\right),
\end{equation}
where $F$ is an unspecified scaling function. Expansion of $M_{nl}$ in powers of the external uniform field $H$ gives
\begin{equation}
M_{nl}=-b_{3}t^{-\gamma}H^{3}+b_{5}t^{-(2\gamma+\beta)}H^{5}-b_{7}t^{-(3\gamma+2\beta)}H^{7}+...,
\end{equation} Comparing the Eq. (7) with Eq. (5), one can see that the leading nonlinear terms $\chi_{3}$ and $\chi_{5}$ diverge as $T\rightarrow T_{f}$ according to $t^{-\gamma}$ and $t^{-(2\gamma+\beta)}$, respectively. Figure 4 shows the experimentally observed divergence of the coefficients of the first two nonlinear terms in the expansion of the magnetization (estimated from fit of the magnetization following Eq. (4)) as $t \rightarrow 0$ from above in the La$_{0.9}$Sr$_{0.1}$CoO$_3$ single crystal. To obtain the critical exponents $\beta$ and $\gamma$ we have used the scaling laws for the nonlinear magnetization given by Eq. (6). According to this equation, for the proper choice of $\beta$ and $\gamma$, $M_{nl}/t^{(\gamma+3\beta)/2}$ data in the vicinity of $T_{f}$ should fall on a single curve when plotted against $H/t^{(\gamma+\beta)/2}$.
\begin{figure}
  \includegraphics[width=0.5\textwidth]{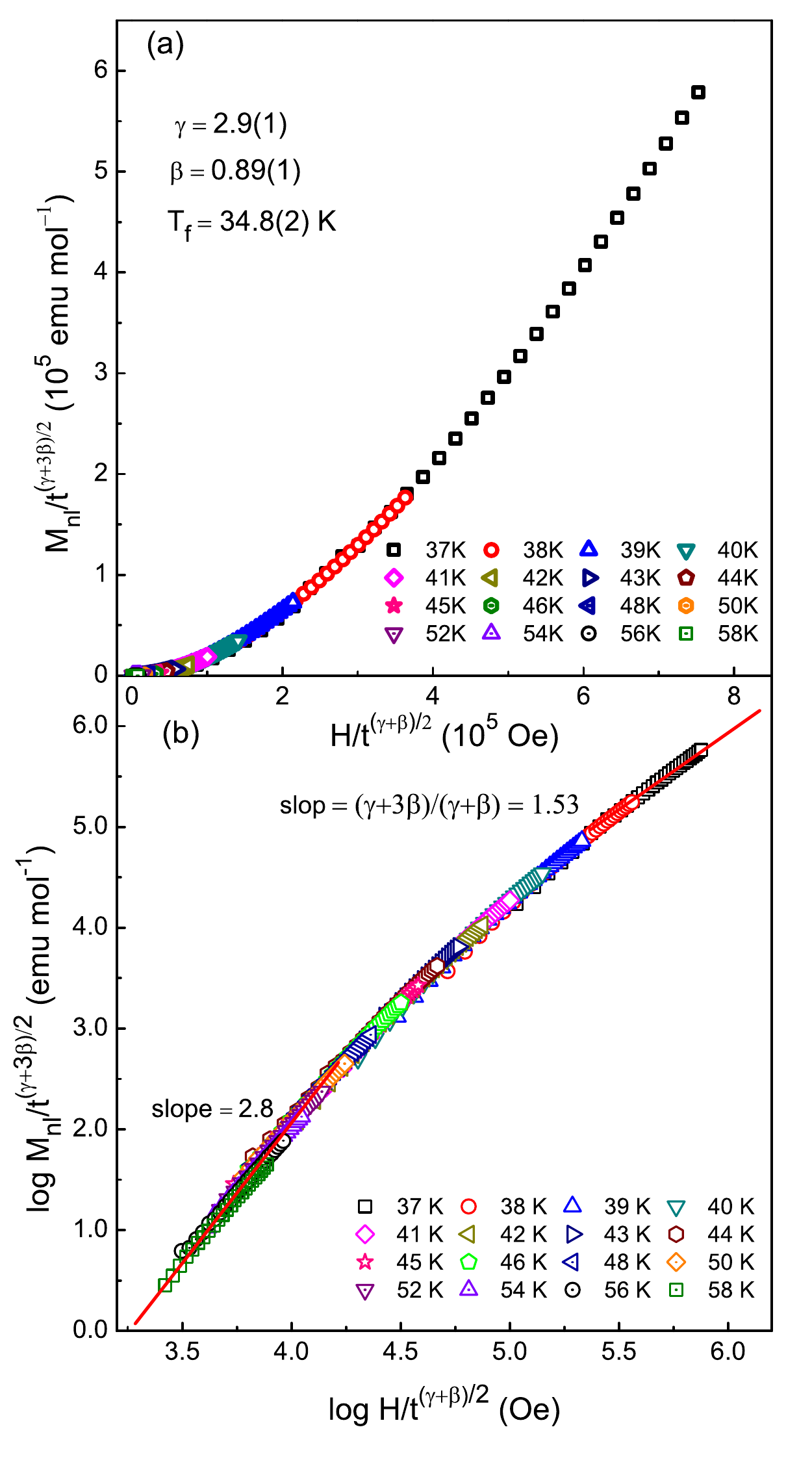}\\
\caption{(Color online) Isothermal magnetization (\emph{M} vs \emph{H}) curves at temperatures in the vicinity of $T_{f}$ in the field range
0$\leq$$H$$\leq$3500 Oe for the La$_{0.9}$Sr$_{0.1}$CoO$_3$ single crystal. The solid lines are due to the fit to Eq. (4).}\label{FIG. 5}
\end{figure}
The best data collapse obtained using an iteration method yields $\beta=0.89(1)$ and $\gamma$=2.9(1) with $T_{f}$=34.8(2) for the present compound and is shown in the Figure 5(a). Figure 5(b) shows the same on a log-log plot with the asymptotic limits of the scaling function. In the region of small nonlinearities i.e., for $T$ significantly larger than $T_{f}$ where the expansion in Eq. (7) retains only the first term, the slope of the scaling curve tends to 3 and for $T$ very close to $T_{f}$ the slope of the scaling curve tends to the full asymptotic value $(\gamma+3\beta)/(\gamma+\beta)$.\cite{mauger} We found a slope of 2.8 from a linear fit of the scaled data for $t\geq0.6$ and a linear fit of the data close to transition($t\leq0.10$) gives a slope of 1.53, which is consistent with the values of $\beta$ and $\gamma$ estimated from the scaling (Figure 5(b)) and therefore implies that these values of $\beta$ and $\gamma$ are reliable and intrinsic to the system. A value of $\gamma$=2.3(1) has also been estimated from the divergence behavior of $\chi_{3}$ ($\chi_{3}$$\propto$$t^{-\gamma}$) through a linear fit on a log-log plot shown in the inset b of Figure 4. Though this value is close to that estimated from the scaling, we have taken the value of $\gamma$ obtained from the scaling as it gives the better data collapse. Using the values of $\beta$ and $\gamma$ estimated experimentally, the other static critical exponents viz., $\delta$, $\eta$, and $\nu$ have been estimated using scaling and hyperscaling relations for a proper comparison between experiment and theory and displayed in Table I. The values of $\delta$, $\nu$, and $\eta$ are obtained from scaling and hyperscaling relations $\delta=1+(\gamma/\beta)$, $d\nu=2\beta+\gamma$ with the dimensionality $d$=3 for the present system, and $\eta=2-(\gamma/\nu)$, respectively. The estimated values of different exponents for the La$_{0.9}$Sr$_{0.1}$CoO$_3$ single crystal fall well inside the realm of known experimental 3D Heisenberg systems\cite{lago,mauger,bitla,bouchiat,plevy} and matches quite well with that of the well known canonical spin glass AgMn\cite{plevy} that further establishes a true low temperature spin glass phase in the present compound. The recent theoretical models which give better estimates of different critical exponents are 3D bimodal ($\pm J$) or Gaussian Heisenberg chiral spin glasses (HCSGs) with weak random anisotropy\cite{viet,kawamura,hukushima} and 3D bimodal or Gaussian Ising spin glass system\cite{hasenbusch,katzgraber} and their results are shown in Table I. Besides the widely different values of the exponents for the HCSGs and ISGs, the opposite sign of $\eta$ and the large difference between the values of $\gamma$ clearly distinguish the two universality classes.  For the present compound we found that $\eta$ is positive and the value of $\gamma$ is much smaller than that of the ($\pm J$) Ising spin glass (ISG) model, which imply that the present system belongs to the 3D Heisenberg universality class. It is worthy to mention that the estimated values of the exponents for the La$_{0.9}$Sr$_{0.1}$CoO$_3$ single crystal are also in consistent with the recent experimental studies in the related systems.\cite{lago,bitla} \\
\begin{table*} {
\caption{Comparison of the deduced critical exponents of La$_{0.9}$Sr$_{0.1}$CoO$_3$ single crystal with canonical spin glass AgMn and different theoretical
models viz., the bimodal ($\pm J$) Heisenberg chiral spin glass (HCSG), the Gaussian (GHCSG), the ($\pm J$) Ising spin glass (ISG), and the GISG. }
\label{I}
\begin{tabular*}{1.0\textwidth}{@{\extracolsep{\fill}}c c c c c c c }
\hline 
 Exponent & AgMn (Ref.\cite{plevy}) & LSCO ($x=0.1$) & $\pm J$ HCSG (Ref.\cite{hukushima}) & GHCSG (Ref.\cite{viet,kawamura}) & $\pm J$ ISG
 (Ref.\cite{hasenbusch}) & GISG (Ref.\cite{katzgraber})\\
\hline
$\beta$ & 0.9(2) & 0.89(1) & 1.2(7) & 1.1(3) & 0.77(5) & 0.77(5) \\[6pt]
$\gamma$ & 2.3(2) & 2.9(1) & 1.5(4) & 2.0(5) & 5.8(4) & 5.8(3) \\[6pt]
$\delta$ &3.3(3) & 4.2(2) & 2.3(4) & 2.75(4) & 8.6(1) & 8.5(8) \\[6pt]
$\eta$ & 0.23(32) & 0.14(11) & 0.8(2) & 0.6(2) & -0.375(10) & -0.37(5) \\[6pt]
$\nu$ & 1.30(15) & 1.56(4) & 1.2(2) & 1.4(2) & 2.45(15) & 2.44(9) \\[6pt]
$z$ & 5.3(8) & 6.1(3) &  &  &  &  \\[6pt]
\hline
\hline
\end{tabular*}}
\end{table*}

\subsection{H$-$T phase diagram}
Experimental studies on some Heisenberg spin glass systems revealed two irreversibility regimes for a fixed uniform magnetic field which are dictated by a weak (higher temperatures) and a strong (lower temperatures) irreversibility lines, $T_{s}(H)$ and $T_{w}(H)$, respectively, in the $H-T$ phase diagram.\cite{kenning} The presence of these two irreversibility lines is also predicted by the existing theoretical models. The infinite-range Sherrington-Kirkpatrik model\cite{sherrington} predicts a finite-temperature ($T_{AT}(H)>0$) phase transition in the absence and presence of external magnetic fields for both Ising and Heisenberg spin glass systems and the evolution of $T_{AT}(H)$ with the external field $H$ is governed by the d'Almeida-Thouless (AT) phase transition line of the following form,\cite{almeida}
\begin{equation}
t_{AT}^3=[1-(T_{AT}(H)/T_{AT}(0))]^3=(3/4)h^2,
\end{equation}
where $h=g\mu_{B}H/k_{B}T_{AT}(0)$, $g$ being the Lande $g$-factor. Experimentally, on the low-temperature side of the AT line there is a phase with broken replica symmetry, while on the high-temperature or high-field side there is a replica symmetric paramagnetic state. However, according to the droplet scaling theory,\cite{fisher} there should be no AT line i.e., no true phase transition in presence of an external field, as in the case of a ferromagnet where the addition of a field removes the phase transition. In this model, the low-temperature phase in zero field is replica symmetric. Gabay and Toulouse (GT) extended the calculation for the SK model for classical isotropic $n$-component vector spin glass and predicted two successive field-dependent transitions; the former is associated with the freezing of transverse spin component and the later one is associated with the longitudinal spin component.\cite{gabay} The freezing of the transverse spin component which is associated with the onset of weak irreversibility occurs along the so-called GT line, a true transition line governed by
\begin{equation}
t_{GT}=1-(T_{w}(H)/T_{w}(0))=[(n^2+4n+2)/4(n+2)^2]h^2,
\end{equation}
The freezing of the longitudinal spin components takes place along a second transition line,
\begin{equation}
t_{AT^{'}}^3=[1-(T_{s}(H)/T_{s}(0))]^3=[(n+1)(n+2)/8]h^2,
\end{equation}
which is associated with the onset of strong irreversibility of the frozen state. \\
The presence of a weak random anisotropy due to the Dzyaloshinsky-Moriya (DM) or dipolar interaction and uniaxial anisotropy fields significantly modify the above picture.\cite{vieira,kotliar} The influence of random DM interaction on the phase transition in 3D Heisenberg systems has been investigated theoretically, which predicts a crossover from Ising to isotropic-like or from a weak to a strong irreversibility behavior with increasing applied magnetic field.\cite{kotliar} Such irreversibility crossover behavior due to the presence of a weak random anisotropy has been experimentally observed in some real spin glasses like CuMn alloy, Cd$_{0.62}$Mn$_{0.38}$Te, and SrFe$_{0.9}$Co$_{0.1}$O$_{3.0}$.\cite{kenning,lago} Another theoretical model of classical $n$-component vector spin glass, which considers the presence of uniaxial anisotropy fields, predicted a new feature, in particular, the reentrant behavior of the GT line in the $H-T$ phase diagram.\cite{vieira} According to this model,\cite{vieira} for $D,H$$\ll$$J$, where $D$ and $J$ being the mean strength of the anisotropy and exchange coupling, respectively, the so-called GT line exhibits a linear behavior governed by $t_{GT}$$\sim$ $-H/J$, for $H\ll D$; however, if $H\gg D$ one observes a crossover to the usual GT behavior, $t_{GT}$$\sim$ $(H/J)^{2}$, predicted by Eq. (9). Such reentrant effect has been observed in a heavy fermion spin glass URh$_{2}$Ge$_{2}$ where the reentrant behavior is associated with the GT line exhibiting a decrease in the freezing temperature with decreasing magnetic field.\cite{maksimov} Similar field dependence of the freezing temperature has also been observed in different other real spin glasses.\cite{lima,bbarbara,lundgren,barbara,keener} \\

\begin{figure}
  \includegraphics[width=0.5\textwidth]{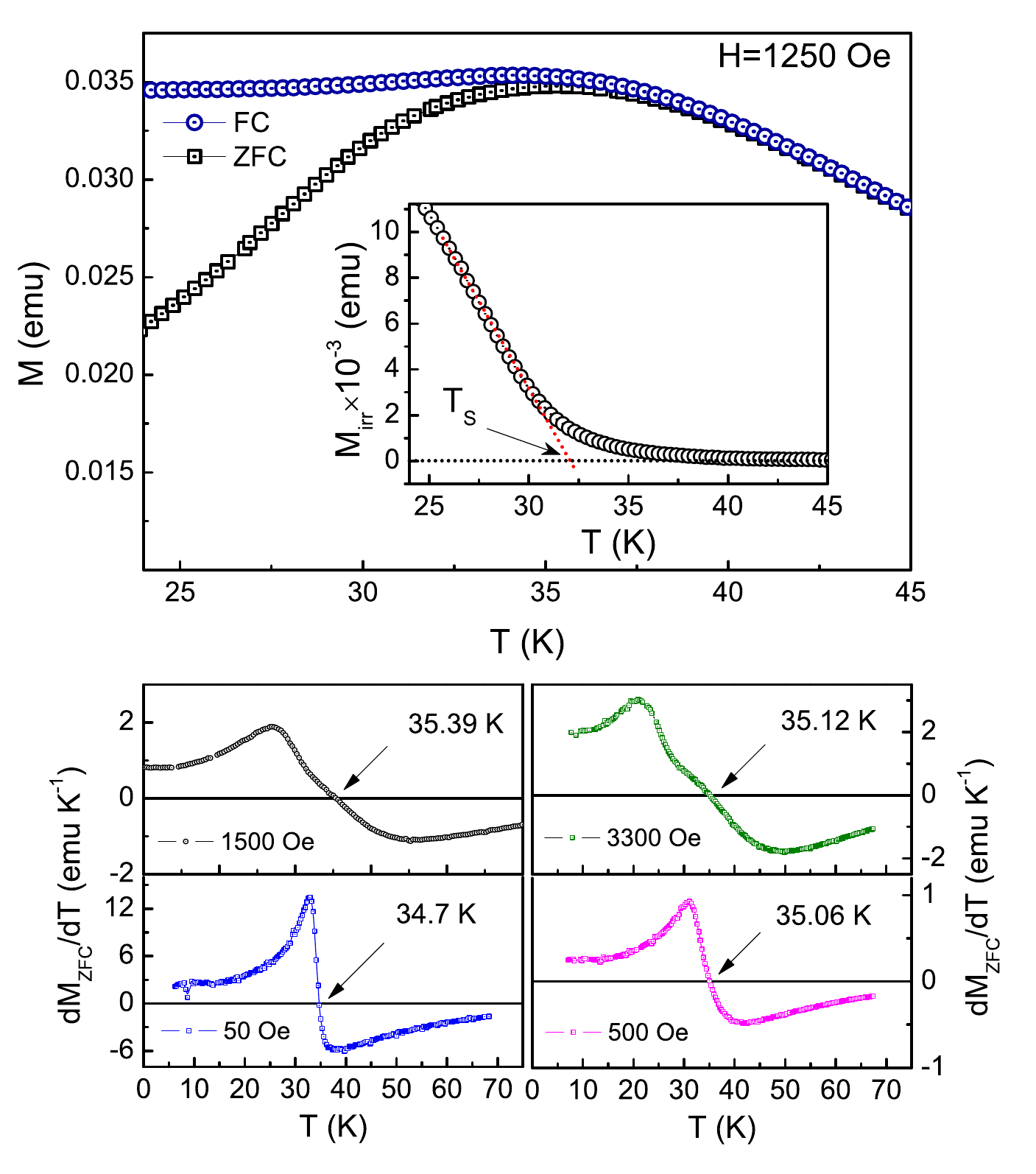}\\
\caption{(Color online) (Top) Temperature dependence of  ZFC and FC magnetization at $H_{appl}=$1250 Oe. Inset shows the procedure to estimate strong irreversibility temperatures from the irreversible magnetization ($M_{irr}$). (Bottom) Different panels show the temperature derivative of the ZFC magnetization at different applied uniform fields. Weak irreversibility temperatures at different fields are estimated from the temperature where $dM_{ZFC}/dT$ becomes equal to zero.}\label{FIG. 6}
\end{figure}

The top panel of Figure 6 shows the ZFC and FC magnetization curves in an applied field of 1250 Oe. Relative to the ZFC curve at 50 Oe in Figure 1, the cusp in the ZFC at 1250 Oe shows a significant broadening with the transition temperature shifted towards the high temperature as found from its temperature derivative ($dM_{ZFC}/dT=0$). The ZFC curve at 1250 Oe reveals that with decreasing temperature it first attains a maximum at the so called freezing temperature $T_{f}(H)$=35.32 K, nearly the same temperature at which the FC and ZFC start to bifurcate and then show a sudden down turn occurring at a lower temperature $T_{s}(H)$ which marks the onset of strong irreversibility. At low field $T_{s}(H)$ and $T_{f}(H)$ almost coincide. Similar behavior has been observed in different real spin glass systems including the recently studied SrFe$_{0.9}$Co$_{0.1}$O$_{3.0}$ compound by Lago $\emph{et al.}$\cite{lago} \\
\begin{figure}
  \includegraphics[width=0.5\textwidth]{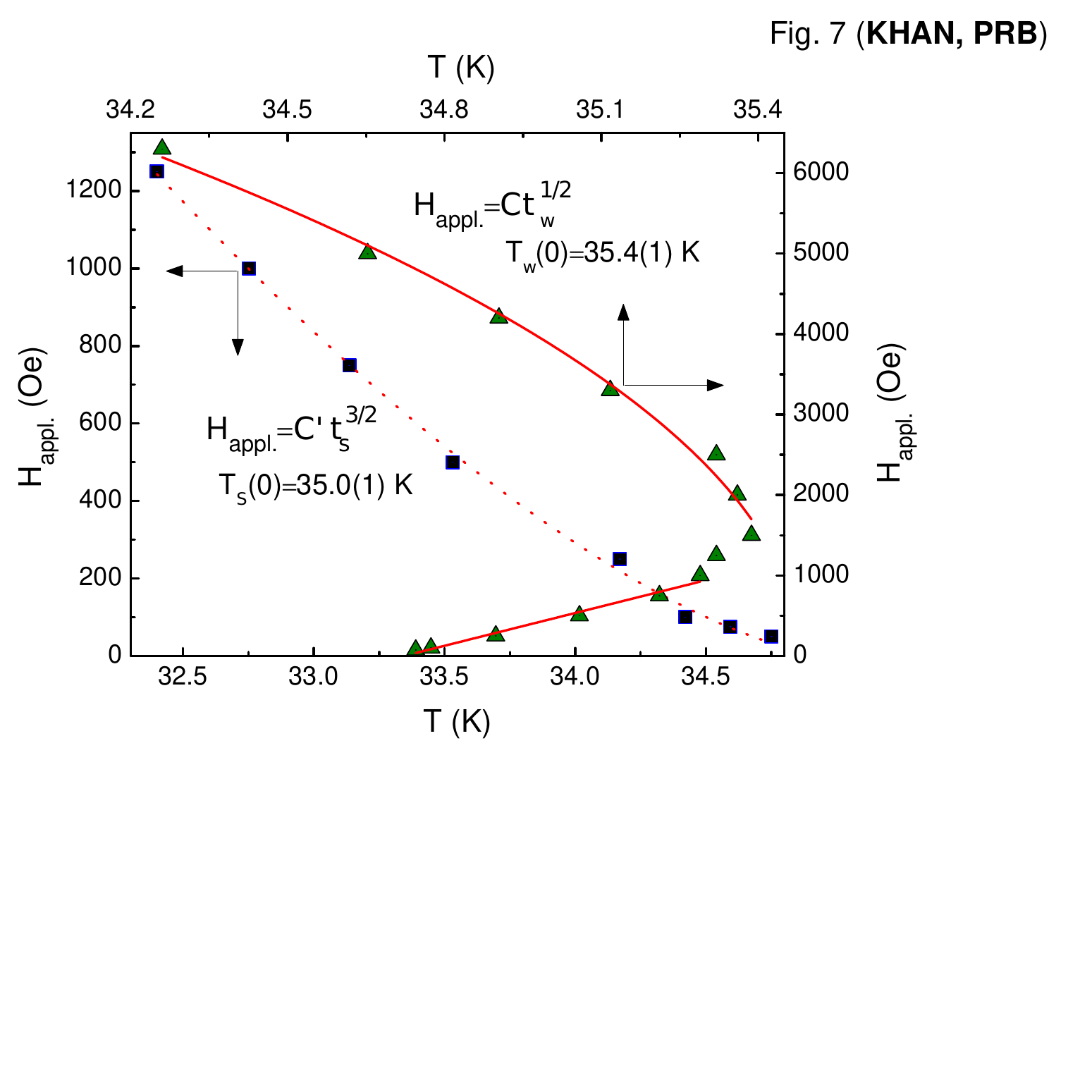}\\
\caption{(Color online)  Magnetic phase diagram of La$_{0.9}$Sr$_{0.1}$CoO$_3$ in the $H-T$ plane. The dotted  line is the fit of the strong irreversibility temperatures to the d'Almeida-Thouless, $H=C^{'}t_{s}^{3/2}$, critical line. The solid lines are the linear fit and the Gabay-Toulouse, $H=C t_{w}^{1/2}$, critical line fit for $H<1000$ Oe and 1500$<$$H$$<$6500 Oe, respectively.}\label{FIG. 7}
\end{figure}

The $H-T$ phase diagram of the La$_{0.9}$Sr$_{0.1}$CoO$_3$ single crystal is shown in Figure 7, which is obtained from the dc magnetization measurements performed using standard procedure. The onset of freezing in an uniform external field $H$ can be ascribed to the weak irreversibility temperature $T_{w}(H)$, which is obtained from the temperature derivative of the ZFC magnetization ($dM_{ZFC}/dT=0$) at that field as shown in the lower panels of Figure 6.\cite{chamberlin,maksimov} The onset of strong irreversibility $T_{s}(H)$ is obtained from irreversible magnetization measurements following standard procedure of extrapolating the linear part of $M_{irr}=M_{FC}-M_{ZFC}$ in the strong irreversibility region to zero as can be seen from the inset of the top
panel of Figure 6.\cite{kenning,lago} The field dependence of strong irreversibility temperature is fitted to an AT like line $H_{appl.}=C^{'}t_{s}^{3/2}$ which gives zero field transition temperature $T_{s}(0)$=35.0(1) K which is close to the freezing temperature $T_{f}$=34.8(2) K obtained from the scaling analysis. The strong irreversibility behavior in the present case is qualitatively similar to that observed in 3D Heisenberg spin glass systems.\cite{kenning,lago} However,
the weak irreversibility temperatures $T_{w}(H)$ in the plane of magnetic field versus temperature shows a reentrant behavior where the $T_{w}(H)$ first increases with increasing applied field, attains a maximum at $H$$\sim$1500 Oe and then decreases following the so-called GT line as shown in the Figure 7. For $H<1000$ Oe, the GT line shows a linear behavior with a negative slope and for $H$$>$1500 Oe a crossover to the usual GT behavior is observed.  The Gabay-Toulouse power law $H_{appl.}=Ct_{w}^{1/2}$ can be fitted for applied field in the range 1500$<$$H$$<$6500 Oe, and yields zero field transition temperature $T_{w}(0)$=35.4(1) K. Qualitatively, the reentrant behavior of the GT line for the present compound is in good agreement with that predicted by the theoretical model for Heisenberg spin glass with a uniaxial anisotropy as discussed earlier.\cite{vieira} A reentrant behavior of the GT line has also been observed experimentally in a heavy fermion URh$_{2}$Ge$_{2}$ Ising spin glass\cite{maksimov} and other Heisenberg spin glasses\cite{barbara,keener} with uniaxial anisotropy fields. Therefore, the observed reentrant behavior of the GT line in the La$_{0.9}$Sr$_{0.1}$CoO$_3$ single crystal is attributed to the presence of a finite single-ion anisotropy in the compound. So, the obtained $H-T$ phase diagram for the present compound is consistent with the 3D Heisenberg spin glass behavior. It should be mentioned that the behavior of the GT line can be further validated by means of more appropriate techniques like the torque measurements which give the direct information about the transverse magnetization.\cite{lago}

\section{conclusion}

We have presented a comprehensive study on the spin glass behavior in La$_{0.79}$Sr$_{0.21}$CoO$_{3}$ single crystal by ac susceptibility and dc magnetization
measurements. The study confirms that the compound undergoes a true equilibrium spin-glass transition at about 34.8(2) K. The behaviors of ac susceptibility are the characteristics of a canonical spin glass and are not due to the dynamical blocking of spin clusters though the system exhibits magnetoelectronic phase separation. The analysis of the static critical behavior through scaling hypothesis yields a series of critical exponents which fall well within the realm of known 3D Heisenberg canonical spin glasses. The obtained $H-T$ phase diagram also supports qualitatively the 3D Heisenberg spin glass behavior and predicts the presence of a uniaxial anisotropy field in this compound. Our study reveals that the La$_{0.79}$Sr$_{0.21}$CoO$_{3}$ single crystal exhibits 3D Heisenberg spin glass behavior in an insulating state where such behavior was previously observed only in a handful systems like SrFe$_{0.9}$Co$_{0.1}$O$_{3}$,\cite{lago} CdCr$_{2}$InS,\cite{kenning} and CdMnTe.\cite{kenning,mauger} It may be mentioned that to our knowledge, this is the only report studying the spin glass critical behavior in a single-crystalline material which gives the intrinsic properties. Further studies in this direction with other doping levels approaching the percolation where the FM cluster effect dominates may be performed for understanding the concomitant disappearance of the SG state and the evolution of a long-ranged FM state in LSCO.        \\

\section{Acknowledgement}

The authors would like to thank A. Pal and D. Bhoi for technical assistance.  \\

\end{document}